\documentclass[prl, twocolumn,amsmath,amssymb, floatfix]{revtex4}
\usepackage{graphicx}
\usepackage{color}

\newcommand{\R}{{\mathbf R}}
\newcommand{\rr}{{\mathbf r}}
\newcommand{\F}{{\mathbf F}}
\newcommand{\G}{{\mathbf G}}
\newcommand{\e}{{\mathbf e}}
\newcommand{\HH}{{\mathbf H}}
\newcommand{\uu}{{\mathbf u}}

\begin{document}
\title{Hydrodynamic Interactions in Two Dimensions}
\author{R.~Di Leonardo$^1$}
\email[]{roberto.dileonardo@phys.uniroma1.it}
\author{S.~Keen$^2$}
\author{F.~Ianni$^{1,3}$}
\author{J.~Leach$^2$}
\author{M.~J.~Padgett$^2$}
\author{G.~Ruocco$^{1,3}$}
\affiliation{
$^1$ CNR-INFM, CRS SOFT c/o Dipartimento di Fisica, Universit\'a di Roma ``La Sapienza'', I-00185, Roma, Italy\\
$^2$ SUPA, Department of Physics \&  Astronomy, University of Glasgow, Glasgow, Scotland\\
$^3$ Dipartimento di Fisica, Universit\'a di Roma ``La Sapienza'', I-00185, Roma, Italy\\
}
\date{\today}
\begin{abstract}

\noindent 

We measure hydrodynamic interactions between colloidal particles confined in a
thin sheet of fluid. The reduced dimensionality, compared to a bulk fluid,
increases dramatically the range of couplings. Using optical tweezers we force
a two body system along the eigenmodes of the mobility tensor and find that
eigen-mobilities change logarithmically with particle separation. At a hundred
radii distance, the mobilities for correlated and anti-correlated motions
differ by a factor of two, whereas in bulk fluids, they would be practically
indistinguishable. We derive the two dimensional counterpart of the Oseen
hydrodynamic tensor which quantitatively reproduces the observed behavior.
These results highlight the importance of dimensionality for transport and
interactions in colloidal systems and proteins in biological membranes.

\end{abstract}
\maketitle


When governing a world of lower dimensionality, the laws of physics give rise
to intriguing phenomena. The reduced number of spatial dimensions usually
results in stronger and longer ranged correlations.
One example of this is long range Coulomb correlations and electron-lattice
interactions which give rise to peculiar electronic and structural phase
transitions in system of low dimensionality \cite{anderson}.
In a similar way, fluid flow propagators, mediating hydrodynamic interactions
between suspended bodies, have logarithmic tails in two dimensions
\cite{batchelor}, giving rise to strong dynamical correlations \cite{cheung}.

When two particles suspended in a viscous fluid approach each other,
propagation of fluid flow results in hydrodynamic interactions.  As a
consequence, a particle's motion resulting from a given stimuli, strongly
depend on the entire instantaneous spatial configuration.  Hydrodynamic
interactions tend to favour correlated motions where every particle can move
within the "slip-stream" of its neighbours \cite{rdl}.  This has consequences
on many collective phenomena such as colloidal aggregation and gel formation
where attractive interactions push to modify interparticle distances
\cite{tanaka}. Moreover particles are always subject to stochastic thermal
forces whose effects may be strongly influenced by hydrodynamic interactions,
as happens in polymer dynamics \cite{doi}, or for protein conformational
changes \cite{delatorre, hayward}. In three dimensional bulk fluids, the
strength of hydrodynamic coupling decays as the inverse interparticle distance.
This long range character makes hydrodynamic interactions quite effective in
determining dynamical behavior and poses a number of numerical and theoretical
challenges to the physical modeling of such phenomena.

The situation is even more dramatic when fluid flow is restricted to two
dimensions, such as in cell membranes or soap films, and flow fields decay
logarithmically with distance.  It is important to note that spatially
confining the particles to two dimensions is not sufficient to observe these
long range tails.  It is essential that the momentum flow is similarly
restricted on a two dimensional plane.  To this aim it is crucial that the
bounding fluid has a much smaller viscosity than the film itself, a solid
boundary would lead instead to hydrodynamic interactions decaying faster than
in the 3D case \cite{cui}. 
 
The problem of two dimensional, single particle mobilities has received a lot
of attention due to its relevance for Brownian motion in cellular membranes
\cite{saffman}.  However, much less is known about the equally non trivial and
important role of many body hydrodynamic interactions in two dimensional
systems. Using video microscopy, Cheung and coworkers \cite{cheung} observed
long ranged spatial correlations between spheres floating in a free standing
liquid film.  Manybody effects were found to be significant even at very small
concentrations, practically precluding the possibility of isolating
hydrodynamic effects from indirect bulk measurements.

In this Letter we demonstrate how optical micromanipulation allows the direct
observation of hydrodynamic interactions between an isolated pair of colloidal
particles suspended in a free standing liquid film.  Optical tweezers allow the
positioning of two beads at varying separations, isolated from boundaries and
other particles.  The tweezers can also be used to simultaneously drive
particles along any direction and directly probe the full mobility tensor.  As
a result, we observe a dramatic increase in the interaction range of
hydrodynamic interactions due to the reduced dimensionality. On the other hand,
we derive a two dimensional expression for the Oseen hydrodynamic tensor.  Once
the relevant far field boundary conditions are recognized, we can
quantitatively describe experimental observations. 

We are interested in determining particle motions under the action of an
external force field. To the lowest order in hydrodynamic interaction (large
enough interparticle distances), it is sufficient to know single particle
quantities and in particular the mobility $b$ and the far field
Green's tensor $\G(\rr)$ describing flow propagation:
\begin{eqnarray}
&&\dot\R_i=b\F_i\\
&&\uu(\rr)=\G(\rr-\R_i)\cdot\F_i
\end{eqnarray} 

\noindent where $\uu(\rr)$ is the flow field produced by a force $\F_i$ applied
to a particle located in $\R_i$.

In a many body system the velocity of the i-th particle will be (within the
superposition approximation) the sum of two contributions: the speed that it
would have in the absence of other particles plus an ``ambient" velocity
obtained as the sum of all the fluid velocities independently produced at
$\R_i$ by other particles located at $\R_j$. 

\begin{equation} 
\label{motion}
\dot{\R}_i=b{\F}_i+\sum_{j\neq i}\G(\R_i-\R_j)\cdot\F_j
\end{equation}

In bulk three dimensional fluids, $b$ is the inverse Stokes drag ($1/6\pi\mu a$)
while $\G$ is the Oseen tensor \cite{doi} which, in cartesian coordinates,
reads:

\begin{equation}
\label{propagator3D}
G_{3D}^{\alpha\beta}(\rr)=
\frac{1}{8\pi\mu r}\left[
\delta_{\alpha\beta}
+\frac{r^\alpha r^\beta}{r^2}
\right]
\end{equation}

If we neglect stresses produced by the fluid bounding
the film (air in our case) then the variations of flow properties across the
film are negligible and dynamics is governed by a two dimensional Stokes
equation. In other words, we only allow for in plane momentum flow and neglect
contributions from the film surfaces.  
Working with relatively slow flows at the micron scale, the Reynolds number
will be negligibly small  allowing to express the momentum balance equation in
the form of a two dimensional Stokes equation:
\begin{equation}
\label{stokes}
\mu\nabla^2\uu(\rr)-\nabla p(\rr)=-\F\delta(\rr)/h
\end{equation}

When complemented by the incompressibility condition $\nabla\cdot\uu=0$,
equation (\ref{stokes}) can be rewritten as a biharmonic equation and solved
for the propagator: 

\begin{equation}
\label{propagator}
G_{2D}^{\alpha\beta}(\rr)=
\frac{1}{4\pi\mu h}\left[
\delta_{\alpha\beta}
\left(\log\frac{L}{r}-1\right)
+\frac{r^\alpha r^\beta}{r^2}
\right]
\end{equation}
where we have neglected $1/r^2$ terms, whose coefficients depend on particle
size and shape and boundary conditions on its surface. Such terms soon become
negligible when moving away from the particle and play no role for hydrodynamic
interactions between sufficiently separated particles. However, these terms
determine the actual value of velocity on the particle boundary and therefore
the single particle mobility $b$. The logarithmic term appearing in
(\ref{propagator}) precludes the possibility of determining the integration
constant $L$ by imposing a vanishing velocity at infinity. The divergence of
flow field signals the presence of a length scale beyond which some of the
assumed approximations fails. 

There are three approximations involved in the derivation of Eq.
(\ref{stokes}): 1) infinite size of the film, 2) negligible inertia, 3)
negligible viscous drag on the interfaces. For each of them there is a length
scale beyond which the solution in (\ref{propagator}) is not self consistent
with the assumed approximation.  For the first approximation, the length scale
is clearly the actual size of the film $L_1$. Following \cite{saffman} we can
impose a frictionless boundary on a ring of radius $L_1$ and obtain the
expression (\ref{propagator}) for the propagator where $L=L_1$.  For a stick
boundary condition one gets $L=L_1/\sqrt{e}$ \cite{brenner}.  In the second
case, the flow field propagated by (\ref{propagator}) fails to satisfy the
negligible inertia approximation
when the distance $r$ is of order $L_2=\mu/\rho U$, where $\rho$ is the fluid
density and $U$ is the typical particle speed. However, inertial terms in the
Stokes equation can be approximately taken into account by the Oseen method
\cite{lamb} and obtain a short distance ($r\ll L_2$) expression for the
propagator which is given by (\ref{propagator}) with $L=4
\exp[1-\gamma]L_2\simeq 6.1 L_2$ ($\gamma$ is the Euler-Mascheroni constant).
At last, 
one can obtain from Saffman solution \cite{saffman} that the momentum flow
through film interfaces cannot be neglected for distances of the order of
$L_3=h \mu/\mu'$ where $h$ is the film thickness and $\mu$, $\mu'$ are
respectively the viscosities of the film and that of the bounding fluid.
However, even in that case, expression (\ref{propagator}) remains valid in a
neighborood $r\ll L_3$ when we replace $L$ with $\exp[1/2-\gamma]L_3\simeq 0.9
L_3$ . When dealing with mesoscopic systems however, due to the very low
Reynolds number involved, $L_2$ is usually very large, a few hundred meters in
our case.  $L_3$ is also quite big when the bounding fluid has a much lower
viscosity than the film fluid, as in the present case where $L_3\sim0.1$m.
Therefore, we expect that hydrodynamic interactions in free standing liquid
films are usually dominated by the finite size of the membrane. The following
discussion will remain valid even in other situations, provided one uses the
relevant length scale $L$. 

We can arrange all $\R_i$ in a single $2 N$ dimensional vector and introduce
the $2 N\times 2 N$ hydrodynamic mobility tensor $\HH$ so that (\ref{motion})
reads:

\begin{eqnarray} 
\label{mobilityeq}
&&\dot{\R}=\HH(\R) \cdot \F\\
\label{mobility}
&&H^{\alpha\beta}_{ij}(\R)=\delta_{ij}\delta_{\alpha\beta} b+
(1-\delta_{ij})G^{\alpha\beta}(\R_i-\R_j)
\end{eqnarray}

We now have a two dimensional expression for the propagator (\ref{propagator})
that can be used in (\ref{mobility}) to get the many body mobility tensor as a
function of the two parameters $b$ and $L$.
We will check the validity of our expression for $\HH$ by observing the
dynamical behavior of two colloidal particles confined in a free-standing
liquid film.  In a two body system, if we choose the $x$ axis along the joining
line, $x$ and $y$ dynamics are naturally decoupled for symmetry reasons. We can
then decompose $\HH$ into two $2\times 2$ mobility tensors operating on the
subspaces of $x$ and $y$ coordinates. Any $2\times2$ symmetrical matrix is
diagonal in the coordinate system of the two eigenvectors $(1,1)$ and $(1,-1)$,
corresponding respectively to a rigid translation and a stretching motion.
Without losing any generality we can then assume that our mobility tensor is
fully characterized by its four eigenvalues.
Using expression (\ref{propagator}) in (\ref{mobility}) we can diagonalize the
mobility tensor in the two body case and obtain the four eigenvalues as a
function of interparticle distance $r$:

\begin{eqnarray} \label{mobs}
\lambda_{x\pm}&=&b\left[1\pm\frac{1}{4\pi\mu
h b}\log\frac{L}{r}\right]\\
\lambda_{y\pm}&=&b\left[1\pm\frac{1}{4\pi\mu
h b}\left(\log\frac{L}{r}-1\right)\right] \end{eqnarray}

\begin{figure}[] \includegraphics[width=0.35\textwidth]{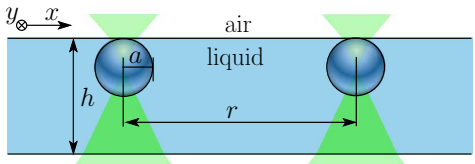}
\caption{(Color online) Trapping geometry. Two beads of radius $a$ are optically trapped
at a distance $r$ in a liquid film of thickness $h$.} 
\label{cartoon} 
\end{figure}

\begin{figure}[ht]
\includegraphics[width=0.5\textwidth]{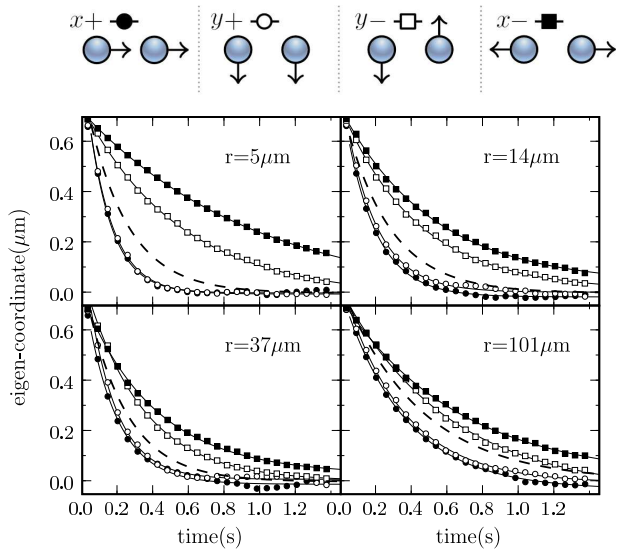}
\caption{Relaxation of eigen-coordinates. For each of the selected distances, we
report the time evolution of the four eigen-coordinates relaxing to equilibrium
after the applied perturbation. Solid lines are exponential fits. Dashed line is
the average, single particle dynamics.}
\label{displacements}
\end{figure}

As in the three dimensional case, hydrodynamic interactions produce a splitting
in the spectrum of mobilities. The splitting is symmetric about the average
single particle mobility and is larger for parallel than for perpendicular
forcing. In three dimensions, the entity of the splitting decays as the inverse
separation falling below 10\% when particles distance grows beyond ten radii.
The dependence on distance, in the two dimensional case, occurs only through a
logarithmic term, which makes hydrodynamic interactions practically
unavoidable.


In order to directly observe 2D hydrodynamic interactions, latex beads (2
$\mu$m diameter) are dispersed in a water-glycerol mixture with 0.2\% wt SDS
surfactant added.  A thin film is obtained by sweeping the solution on a square
frame (6 mm side) of nylon wires (60 $\mu m$ thickness) \cite{capillary}.
Glycerol increases viscosity and slows down both drainage and evaporation,
resulting in longer lived films. Starting with a 50\% wt water/glycerol film
and then heating to evaporate most of the water, we can obtain a very viscous
film with a few micron thickness. We measured the film thickness $h$ at the
beginning and at the end of the reported experiment to be  3.9 $\mu$m.
Particles are imaged by a 40x NA 0.75 objective of an inverted optical
microscope (Nikon TE2000-U).  The same objective is used to focus the laser
beam ($\lambda$=532 nm) diffracted off a spatial light modulator (SLM, Holoeye
LCR-2500) into two, dynamically reconfigurable, optical traps \cite{hot,
optholo}.  Axial confinement is by capillary force on the top surface of the
liquid film (Fig. \ref{cartoon}). 

In contrast to previous studies of hydrodynamic interactions in 3D
\cite{meiners,polin,rdl}, where eigen-mobilities where extracted from the
correlated fluctuations of optically trapped particles, we choose to measure
the eigen-mobilities by directly exciting the four eigenmodes. For each of the
four eigenmodes, we calculate two holograms producing two sets of traps
slightly displaced along the eigencoordinate.  The two holograms are
alternatively displayed onto the SLM. The SLM is fully refreshed with the new
hologram on a time scale (50 ms) which is much faster than the time scale of
particle dynamics. The two particles will be then displaced from the new
equilibrium positions, along the selected eigencoordinate. In our experiment,
external forces are of optical and stochastic origin:

\begin{equation}
\label{force}
\F_i=-k\delta\R_i+{\mathbf S}_i
\end{equation}

\noindent with $k$ the trap strength and $\delta\R_i=\R_i-\R_i^0$ the ith
particle displacement from trap center $\R_i^0$. Assuming small displacements
compared to interparticle distances and averaging over Brownian motion, we get
from (\ref{mobilityeq}, \ref{force}):

\begin{equation} 
\label{eqmot}
\langle\delta\dot{\R}\rangle=-k\HH(\R^0)\cdot\langle\delta\R\rangle
\end{equation}

If the initial configuration corresponds to a small displacement along the nth
eigenmode $\langle\delta\R(0)\rangle=\epsilon \e_n$, then (\ref{eqmot}) has
the solution:

\begin{equation}
\langle\delta\R\rangle=\epsilon \exp[-k\lambda_n t] \e_n
\label{edyn}
\end{equation}

and we can directly obtain the corresponding mobility $\lambda_n$ by monitoring
the amplitude of the $\e_n$ mode relaxing to equilibrium. We choose to
normalize the eigenvectors such that the corresponding eigencoordinates give the
center of mass position, for the rigid modes, and the half distance (along $x$
or $y$) for stretching modes. 

We choose ten logarithmically spaced interparticle separations between 5 $\mu$m
and 100 $\mu$m and drive the two particles back and forth (eight times) along
each of the four eigenmodes.  Particle coordinates where digitally extracted
from video frames at 144 Hz.  Eigencoordinates where then computed and averaged
over the eight iterations.  Fig. \ref{displacements} shows the time evolution of
the four eigen-coordinates at four selected interparticle distances. The two
$x$ and $y$  rigid  motions are much more mobile than the corresponding
stretching motions.  This behavior remains clearly visible up the highest
investigated distance (100 particle radii). The four eigen-mobilities
corresponding to the four probed modes can be extracted by fitting the
eigen-coordinate dynamics to the exponential law in (\ref{edyn}). To correct
for variations in trap strength $k$, for each distance, we normalize the four
obtained decay rates to their average value $k b$.
\begin{figure}[] 
\includegraphics[width=0.45\textwidth]{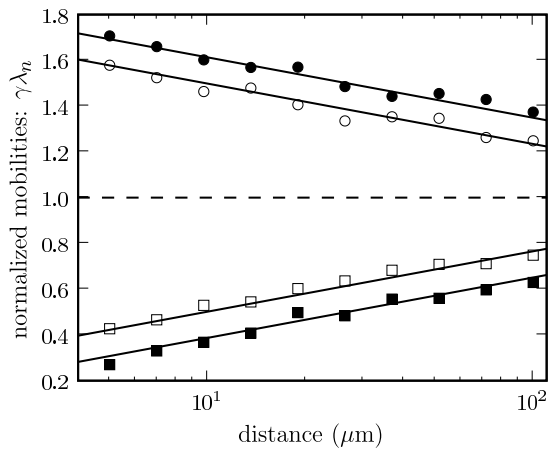}
\caption{Eigen-mobilities. The four eigen-mobilities of a two particle system
arranged at different particles separations. To correct for variations in trapping power,
for each distance mobilities have been normalized to their average.} 
\label{mobilities} 
\end{figure}
The relative eigen mobilities are shown in Fig. \ref{mobilities} as a function
of the particle separation. The strength of hydrodynamic coupling, reflected in
the splitting of mobilities, decays logarithmically slow with distance.  Still
at a separation of 100 radii , particles move twice as fast when forced along the
same direction rather than in the opposite. At the same large separation, three
dimensional mobilities would only differ by 1\%.  

The four data sets in Fig. \ref{mobilities} can be very well fitted by
(\ref{mobs}) leaving $L$ and the adimensional mobility $b^*=4\pi\mu h b$ as the
only free parameters.  We obtain as best fit parameters $b^*=$8.7 and $L=2.1$
mm, the corresponding fitting curves are shown as straight lines.  As expected,
the relevant length scale $L$ is determined by the film finite size. In
particular a sticky boundary condition on a ring inscribed in the film frame
($L_1=3$ mm) would give an $L=1.8$ mm. For the same boundary condition, the
single particle mobility $b$ can be calculated for a cylinder of height $h$,
the film thickness \cite{brenner}.  Using our particle radius $a$ as the
cylinder radius we obtain $b^*=\log(L_1/a)-1=7.0$ which compares reasonably
well with the corresponding fitted value.  We do not expect the two values to
be in better agreement since the mobility $b$ depends on the details of the
boundary conditions on the particle surface.  We anticipate that, when moving
to multiparticle systems, the mobility of long wavelength eigenmodes will
diverge linearly with the number of particles, rather than logarithmically as
in the 3D case.  As a first consequence of that, we could transport, at the
same speed, any number of beads, using the same total amount of laser power.
Moreover, the crossover to underdamped propagating modes on a linear chain of
trapped particles predicted in \cite{polin}, could be experimentally verified
much more easily.

We have directly measured hydrodynamic interactions between colloidal particles
in a sheet of viscous fluid. The reduced dimensionality, compared to the bulk
3D fluids, results in stronger and longer ranged hydrodynamic couplings which
are quantitatively reproduced by a two dimensional version of Oseen
hydrodynamic tensor. The observed interactions constitute a general model for
diffusion and interactions of proteins in biological membranes.



\end{document}